# Chirality-induced quantum nonreciprocity

Zimo Zhang[1,11], Zhongxiao Xu[1,2,11], Ran Huang[3,11], Xingda Lu[4], Fengbo Zhang[1], Donghao Li[1], Şahin K. Özdemir[5], Franco Nori[3,6], Han Bao[7], Yanhong Xiao[2,9], Bing Chen[8,*], Hui Jing[10,*], and Heng Shen[1,2,*]

[1]State Key Laboratory of Quantum Optics and Quantum Optics Devices, Institute of Opto-electronics, Shanxi University, Taiyuan, Shanxi 030006, China

[2]Collaborative research center on Quantum optics and extreme optics, Shanxi University, Taiyuan, Shanxi 030006, China

[3]Quantum Information Physics Theory Research Team, Center for Quantum Computing (RQC), RIKEN, Wakoshi, Saitama 351-0198, Japan

[4]Department of Physics, State Key Laboratory of Surface Physics and Key Laboratory of Micro and Nano Photonic Structures (Ministry of Education), Fudan University, Shanghai 200433, China

[5]Department of Electrical and Computer Engineering, Saint Louis University, St. Louis, Missouri 63103, USA

[6]Physics Department, The University of Michigan, Ann Arbor, Michigan 48109-1040, USA

[7]QUANTUM, Johannes Gutenberg-Universität Mainz, 55128 Mainz, Germany

[8]School of Physics, Hefei University of Technology, Hefei, Anhui 230009, China

[9]State Key Laboratory of Quantum Optics and Quantum Optics Devices, Institute of Laser Spectroscopy, Shanxi University, Taiyuan, Shanxi 030006, China

[10]Key Laboratory of Low-Dimensional Quantum Structures and Quantum Control of Ministry of Education, Department of Physics and Synergetic Innovation Center for Quantum Effects and Applications, Hunan Normal University, Changsha 410081, China

[11]These authors contributed equally: Zimo Zhang, Zhongxiao Xu, Ran Huang.

*e-mail: hengshen@sxu.edu.cn; jinghui@hunnu.edu.cn; bingchenphysics@hfut.edu.cn.

**Chirality, nonreciprocity, and quantum correlations are at the center of a wide range of intriguing effects and applications across natural sciences and emerging quantum technologies. However, the direct link combining these three essential concepts has remained unknown till now. Here, we establish a chiral non-Hermitian platform with flying atoms and demonstrate chirality-induced nonreciprocal bipartite quantum correlations between two channels: Quantum correlation emerges when two spatially separated light beams of the same polarization propagate in opposite directions in the atomic cloud, and it becomes zero when they travel in the same direction. Thus, just by flipping the propagation direction of one of the beams while keeping its polarization the same as the other beam, we can create or annihilate quantum correlations between two channels. We also show that this nonreciprocal quantum correlation can be extended to multi-color sidebands with Floquet engineering. Our findings may pave the road for realizing one-way quantum effects, such as nonreciprocal squeezing or entanglement, with a variety of chiral devices, for the emerging applications of e.g., directional quantum network or nonreciprocal quantum metrology.**

Chirality, the asymmetry of an object and its mirror image, widely exists in nature and plays a key role in essential laws of physics, chemical reactions[1], biological structures[2] and material engineering[3,4], as well as the distribution of galaxies. In quantum physics, chirality provides a powerful tool to control light-matter interactions[5,6,7] towards the realizations of chiral quantum networks, chiral imaging, and directional photonics[8]. By leveraging chiral behaviors[9-16], quantum routers[17], circulators[18], and diodes[19,20], have been realized, offering tools for directional signal processing, back-action-immune communication, and invisible sensing[21]. Yet considerable previous efforts have been devoted to one-way control of coherent light or single photons[5,17-21]. To go beyond what has already been demonstrated, it is essential to find nonreciprocal quantum effects without any classical counterpart, such as one-way anti-bunching or entanglement[22,23]. Such an effort, in turn, requires the control of quantum correlations in a highly directional way. Along this line, nonreciprocity in the correlations of photon pairs in a reciprocal resonator was recently reported[24]. However, quantum nonreciprocity of multipartite correlations, fully induced and controlled by chirality, has not been observed so far.

Here we report the first experiment on chirality-induced breaking of reciprocity in quantum correlations. Specifically, we build a non-Hermitian light-atom system and find clear evidence of quantum nonreciprocity for optical two-channel correlations in the system, i.e., the appearance or the vanishing of quantum correlations by flipping only the flow direction of a laser through the same atoms. We emphasize that the chirality of the light serves as the source of the observed quantum nonreciprocity. In fact, due to chiral light-atom interactions, an effective one-way nonlinearity emerges only for a specific input port while absent for the reversed input port even when the polarization of the light is kept the same. The resulting one-way quantum correlations can be further extended into multi-color quantum nonreciprocities emerging at different frequencies, by using the technique of Floquet engineering, i.e., by periodically driving the non-Hermitian chiral system. Our work establishes the link between chirality and quantum nonreciprocity, making it possible to realize and utilize a variety of chirality-enabled directional quantum effects, such as nonreciprocal squeezing or entanglement, for future applications in chiral quantum optics, nonreciprocal quantum engineering, and one-way quantum sensing.

We note that non-Hermitian chiral phononics was reported very recently[25], which explored only one-way energy flow between mechanical modes in classical regime, and did not consider one-way quantum correlations. Non-Hermitian linear couplings have been studied in classical regime too, for realizing anti-Parity-Time symmetry[26] or photon-magnon interference[27], without mentioning either nonreciprocity or quantum correlation. Even in recent works on non-Hermitian quantum correlations[28], the possibility of achieving nonreciprocal quantum effects and their link to chirality remained unexplored. We also note that classical nonreciprocity was reported in a recent experiment as one-way inter-channel light transport[29], but neither quantum nonreciprocity nor the link between chirality and nonreciprocity was demonstrated. Other experiments revealed directional flow of single photons[17] or self-correlations[18] with coherent chiral coupling of spin-momentum-locked light with emitters in nanostructures. Here, in contrast to all previous works, we use optical chirality to achieve a dissipative and propagation-direction-dependent interaction which enables chirality-induced nonreciprocal quantum correlations. Our work uncovers the first direct connection between two fundamentally distinct asymmetries—chirality and nonreciprocity—at the quantum cross-correlation level. Also, our work highlights the counterintuitive role of dissipation in chiral quantum control, puts forward conceptually new ways for building chirality-controlled one-way quantum devices, and in a broader view, provides a new bridge between a wide range of frontier fields such as chiral quantum optics, nonreciprocal devices, and non-Hermitian physics, as well as Floquet engineering.

Our system consists of two optical channels (CH1, CH2) separated transversely by 1 cm inside an anti-relaxation coated vapor cell containing isotopically enriched $^{87}$Rb vapor with temperature 54 °C. In our experiments, the atoms do not move in a specific direction but instead undergo thermal motion. In the classical nonreciprocity experiments (Fig. 1**a**), the control or probe light couples to the atomic transition of $|1\rangle \to |3\rangle$ or $|2\rangle \to |3\rangle$ with a detuning $\Delta_c$ or $\Delta_p$. In the forward case, with the probe co-propagating with the control beam, an atom always "sees" the probe and control fields with same Doppler frequency shifts, and their effects on the two-photon detuning, $\Delta = \Delta_c - \Delta_p$, are cancelled ($\Delta = 0$), leading to a Λ-type electromagnetically induced transparency (EIT) process. In contrast, for the backward case, when the probe propagates against the direction of the control field, opposite Doppler frequency shifts enhance the two-photon detuning, $\Delta = 2kv$, resulting in the blockade of probe light. Here, $v$ is the velocity of atoms moving along the control field, $k$ is the magnitude of the wavevectors of the control and probe (for simplicity, we take $|k_c| = |k_p| = k$). Such classical nonreciprocity is revealed by the different transmission spectra of the same probe laser for opposite input directions (Extended Data Fig. 2). The mechanism behind this classical nonreciprocity is the Doppler shift induced by the moving atoms, which is independent of the interaction between two channels, as already known in previous works[30, 31].

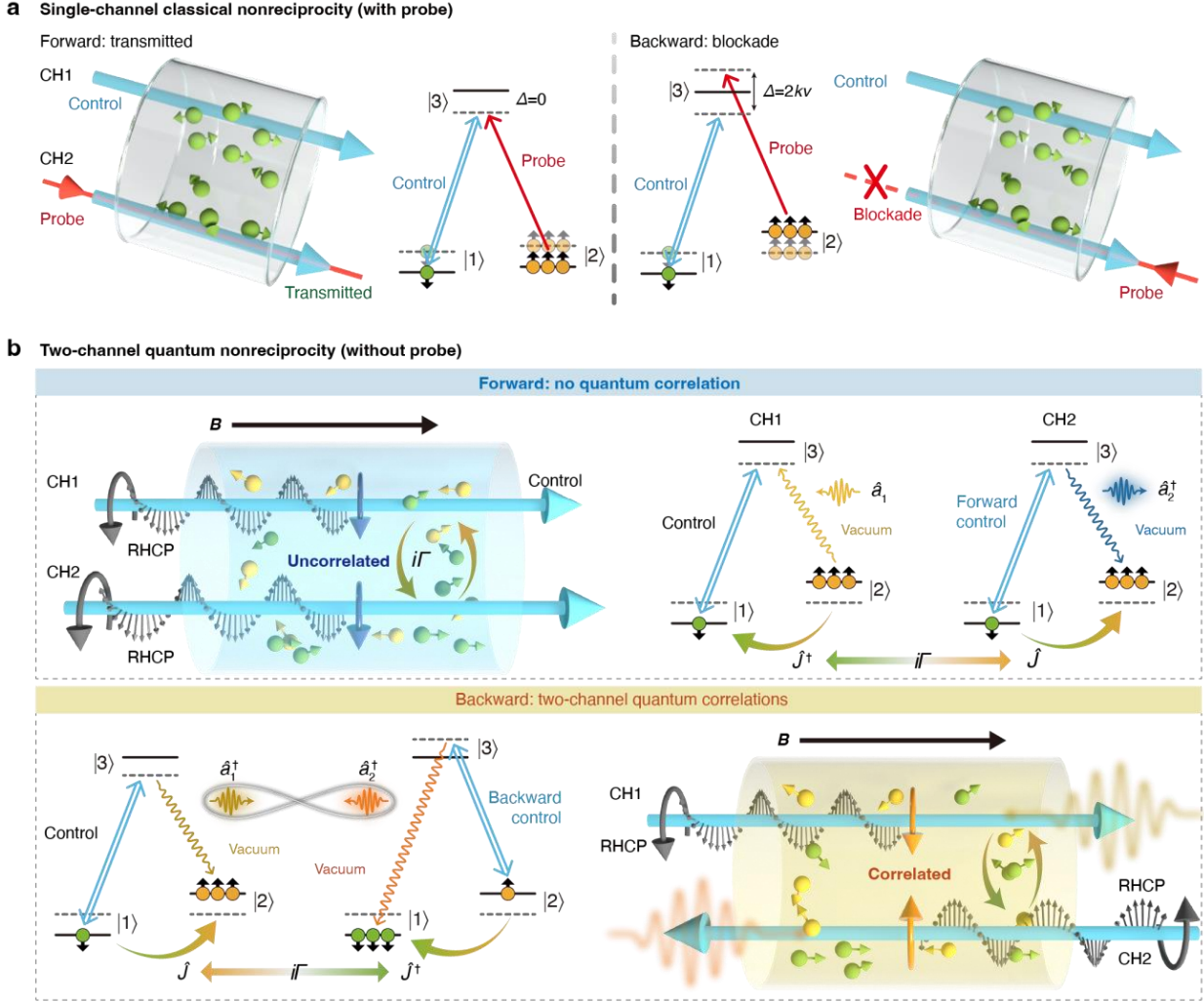

**Fig. 1 | Quantum nonreciprocity versus classical nonreciprocity. a**, Doppler-shift-induced nonreciprocal classical light transmission. Two control lights with the same circular polarizations propagate along the same direction in two channels (CH1 and CH2) with also a probe light propagating along (forward, left) or against (backward, right) the control in CH2. For atoms flying with velocity v along the control, two-photon resonance (detuning), $\Delta = 0$ ($\Delta = 2kv$), occurs in the forward (backward) case due to the same (different) Doppler shifts of the control and probe, resulting in electromagnetically induced transparency, EIT, (blockade of the probe). Here, $k$ is the magnitude of the wavevector. **b**, Chirality-induced quantum nonreciprocity. Without the probe, two control beams with right-hand circular polarizations (RHCP) propagate in the same (forward, upper) or opposite (backward, lower) directions. The coupling of the controls is mediated by the atomic diffusion at a rate $\Gamma$. Quantum nonreciprocity occurs by reversing the input direction of the control in CH2. In the presence of thermal motion, the atoms "see" same (blue) and opposite (orange) circular polarizations for the control beam in the forward or backward case. Such chirality leads to different interactions between the channels, resulting in correlations in the backward case but not in the forward case. The ground states $|1\rangle$ and $|2\rangle$ are Zeeman sublevels of $|F = 2\rangle$, and the excited state $|3\rangle$ is $|F' = 1\rangle$ of the $^{87}$Rb D1 line. A bias magnetic field $\boldsymbol{B}$ is used to shift the homodyne measurement from DC to twice the Larmor frequency ($\sim 300$ kHz) to bypass low-frequency technical noises; $\hat{J}^\dagger$ and $\hat{a}^\dagger_{1,2}$ denote the creations of atomic spin excitation and photons in CH1 (or CH2), respectively.

In quantum nonreciprocity experiments, as shown in Fig. 1**b**, we remove the probe laser and let it be vacuum, i.e., only a control field with fixed right-hand circular polarization is set in each channel, which forms the Λ-type EIT configuration, together with a particular optical mode in vacuum fulfilling the energy and momentum conversations in the interaction. Importantly, the collective spin wave $\rho_{12}$ (ground state coherence) created in the channels is dissipatively coupled to each other through the intrinsic ballistic motion of thermal atoms[26,28]. We find that quantum correlations emerge when the control beam in CH2 propagates against the direction of the control in CH1 (backward case), but do not exist when the CH2 control propagates along the direction of the CH1 control (forward case). Here, the chirality of the control beams serves as the source of the observed nonreciprocity. The underlying principle can be understood as follows: When the right-hand circularly polarized (RHCP) control beams in CH1 and CH2 propagate in the same +z-direction (Fig. 2**a**), atoms "feel" the same electric field and thus "see" the same chirality for the fields in CH1 and CH2. This leads to an effective dissipative coupling between the channels that can be described by the linear dissipative beamsplitter (DBS) model. On the other hand, when the RHCP beams in CH1 and CH2 propagate in opposite directions (e.g., beam in CH1 propagates in +z-direction while the beam in CH2 propagates in −z direction), the atoms "feel" different electric fields and thus "see" opposite chirality for the fields in CH1 and CH2 (i.e., atoms see the RHCP field propagating in −z direction as a left-hand circularly polarized, LHCP, beam propagating in +z-direction, effectively creating the situation in which two beams with opposite chirality propagate in the same +z-direction). This then leads to an effective dissipative coupling between the channels that can be described by the nonlinear non-Hermitian parametric-amplifier (NHPA) model (see Methods for more theoretical details). To summarize, the atoms "see" the same or opposite chirality for the two control beams in the forward or backward cases, respectively, leading to an effective propagation-direction-dependent coupling between CH1 and CH2, which thus results in nonreciprocal quantum correlations.

More specifically, in the forward case, the atoms "see" the same chirality for the beams having the same polarization and propagating in the same direction in CH1 and CH2 (the upper panels of Fig. 1b). In this case, one atomic spin excitation $\hat{J}^\dagger$ ($|2\rangle \to |1\rangle$) in CH1 is accompanied by a lower sideband photon annihilation ($|2\rangle \to |3\rangle$) locally, represented as $\widehat{H}_1 \propto \hat{a}_1 \hat{J}^\dagger + h.c.$ This excitation may diffuse either to the dark region outside the beam into the reservoir resulting in dissipation or to CH2 where it interacts with a light of the same polarization. In the latter case, the photon in the lower sideband ($|3\rangle \to |2\rangle$) is forward scattered along with the annihilation of the same spin excitation $\hat{J}$ ($|1\rangle \to |2\rangle$), described by $\widehat{H}_2 \propto \hat{a}_2^\dagger \hat{J} + h.c.$ Thanks to the collective effect buildup along the propagation direction of light, this two-step interaction results in a linear DBS coupling between CH1 and CH2, which is described by the Hamiltonian $\hat{H}_D \propto \hat{a}_1 \hat{a}_2^\dagger - \hat{a}_1^\dagger \hat{a}_2$, and hence no quantum correlation emerges. In contrast, in the backward case, the atoms "see" opposite chirality for the beams having the same polarization but propagating in opposite directions in CH1 and CH2 (the lower panels of Fig. 1**b**). In this case, a photon in the lower sideband ($|3\rangle \to |2\rangle$) is scattered in CH1 along with the annihilation of a spin excitation $\hat{J}$ ($|1\rangle \to |2\rangle$) captured by the Hamiltonian $\hat{H}_1 \propto \hat{a}_1 \hat{J}^\dagger + h.c.$. When it diffuses to CH2, in the reversed $\Lambda$-type EIT polarization configuration, this annihilation of spin excitation $\hat{J}$ ($|1\rangle \to |2\rangle$) in CH1 is equivalent to a spin excitation $\hat{S}^\dagger$ ($|1\rangle \to |2\rangle$) in CH2. Thus, in CH2 the control beam locally interacts with atoms, which results in the annihilation of a spin excitation $\hat{S}$ accompanied by the upper-sideband photon creation ($|3\rangle \to |1\rangle$) in CH2, as described by $\hat{H}_2' \propto \hat{a}_2^\dagger \hat{S} + h.c. = \hat{a}_2^\dagger \hat{J}^\dagger + h.c.$. Different from the forward case, this two-step interaction with collective dissipative coupling produces a nonlinear interaction, $\hat{H}_N \propto \hat{a}_1^\dagger \hat{a}_2^\dagger - \hat{a}_1 \hat{a}_2$, leading to the buildup of quantum correlation between the light denoted by $\hat{a}_1^\dagger$ and $\hat{a}_2^\dagger$ in the channels (also see Supplementary Information). In a broader view, our approach suggests dissipation engineering as a means of creating quantum correlations and provides a feasible way to explore and utilize quantum nature of non-Hermitian chiral systems[32,33].

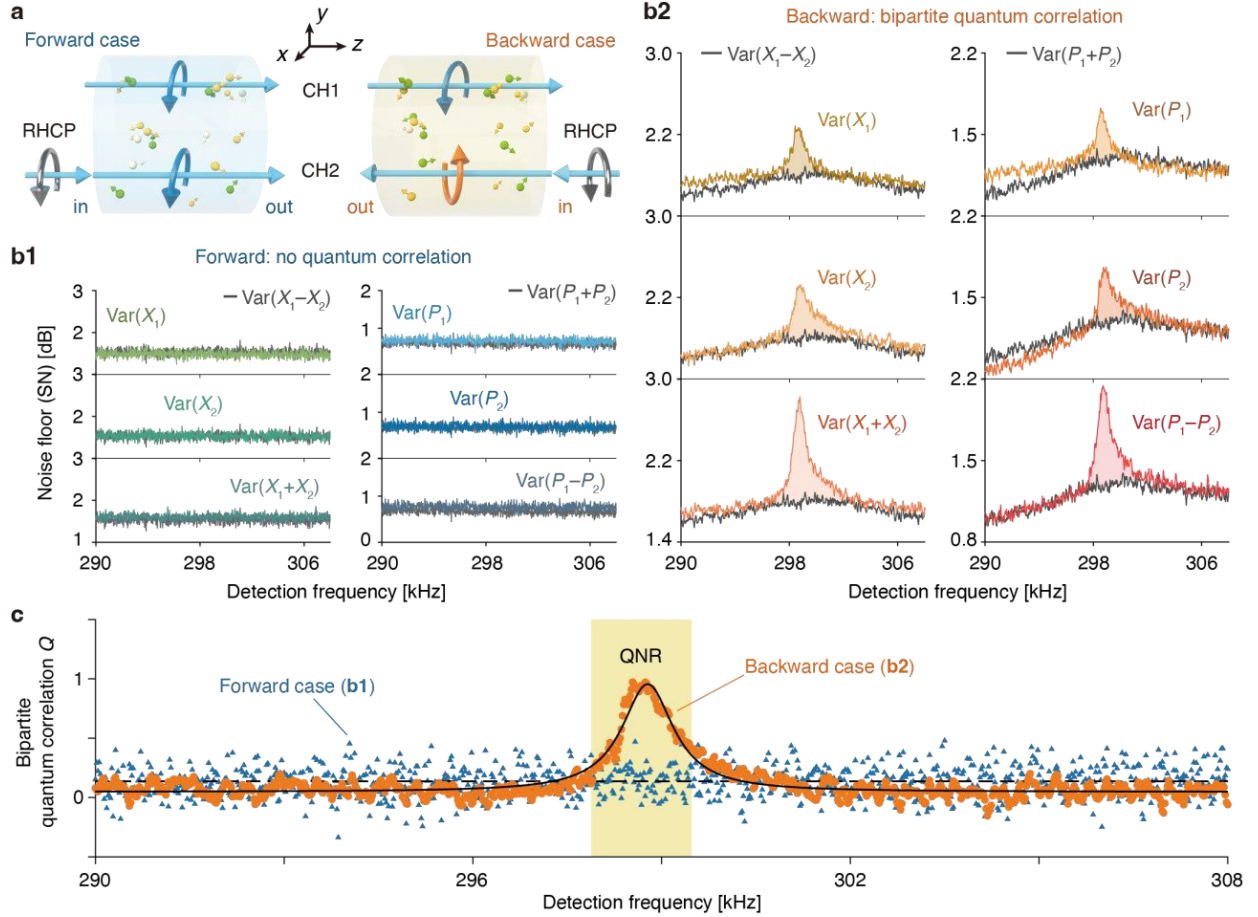

**Fig. 2 | Quantum nonreciprocity of bipartite quantum correlations. a**, Schematics for creating quantum nonreciprocity. Left: two control beams with right-hand circular polarizations (RHCP) propagate along the same direction (forward case). Right: the input direction of the control laser in CH2 is reversed while its polarization is kept as RHCP (backward case). **b**, Measurement results of the quantum noises for the **b1** forward and **b2** backward cases. **c**, Bipartite quantum correlation $Q$. The quantum nonreciprocity (QNR), i.e., bipartite quantum correlation occurs in the backward case (orange) but not for the forward case (blue), is observed around 298.8 kHz (yellow zone). The orange (blue) markers and solid (dashed) curve indicate experimental and theoretical results for backward (forward) case, respectively.

To detect quantum correlations, we apply a bias magnetic field which provides a quantization axis and enables the Zeeman levels in the Λ three-level scheme to be manipulated. In this way, we shift the homodyne detection frequency from DC to twice the Larmor frequency (~ 300 kHz) to bypass the low-frequency technical noise, which then enables optical shot noise-limited measurement of quantum fluctuations. We note that the magnetic field itself does not play any direct role in creating quantum nonreciprocity in our study. With the development of experiment techniques, quantum nonreciprocity may be observed with ultraweak magnetic field[34,35] or even using a fictitious magnetic field induced by light[36].

In our experiment, we use two sets of polarization homodyne detection systems at the output to measure quantum fluctuations of the generated lights in each channel[28], i.e., $\text{Var}(\hat{X}_i)$ and $\text{Var}(\hat{P}_i)$, $(i = 1,2)$ as carried in the photocurrents, where the position and momentum operators of the $i$-th channel are defined as $\hat{X}_i = \hat{S}_x^i/\sqrt{|S_z^i|}$ and $\hat{P}_i = \hat{S}_y^i/\sqrt{|S_z^i|}$, with $\hat{S}_x^i$, $\hat{S}_y^i$, and $\hat{S}_z^i$ denoting the Stokes operators. Here, the control beams act as the local oscillators (for more details, see the Methods). Moreover, in order to evaluate bipartite quantum correlation or quantum correlation between two channels, we perform joint polarization homodyne detection by combining signals from the two sets of polarization homodyne detectors aforementioned into two radio frequency power splitters to obtain the joint variance $\text{Var}(\hat{X}_1 \pm \hat{X}_2)$ and $\text{Var}(\hat{P}_1 \pm \hat{P}_2)$ (see Fig. 2b). Quantum correlation between the fields in CH1 and CH2 can be evaluated from the above measured noise spectra via the formula: $Q = B - A$, with $A = \text{Var}[(\hat{X}_1 - \hat{X}_2)/\sqrt{2}] + \text{Var}[(\hat{P}_1 + \hat{P}_2)/\sqrt{2}]$, and $B = [\text{Var}(\hat{X}_1) + \text{Var}(\hat{X}_2)]/2 + [\text{Var}(\hat{P}_1) + \text{Var}(\hat{P}_2)]/2$. Here, $Q > 0$ ($Q = 0$) indicates the presence (absence) of bipartite quantum correlation. We find that bipartite quantum correlation occurs in the backward case ($Q \sim 0.91$) due to the nonlinear NHPA interaction, while no quantum correlation emerges in the forward case ($Q = 0$) because of the linear DBS interaction (Fig. 2c). The observed quantum nonreciprocity, with $Q \sim 0.91$ of difference between the bipartite quantum correlations for opposite directions, is fundamentally different from the classical nonreciprocity of transmission rates. In addition, we find that when the chirality of the control in CH2 is reversed, such quantum nonreciprocal effect still exists, but the results are reversed, i.e., bipartite quantum correlations occur in the forward case, but not in the backward (Extended Data Fig. 3). We further confirm the direction dependent emergence of quantum correlations using Gaussian discord $\mathfrak{D}_1$—a good indicator of quantum correlations beyond entanglement[37,38]—which has the value of $\mathfrak{D}_1 = 0$ for the forward and $\mathfrak{D}_1 = 2.4 \times 10^{-3}$ for the backward case in our experiment (see Methods and Supplementary Information for more discussions).

The nonreciprocal quantum correlation in our system is restricted to a monochromatic mode (Fig. 2c), because the phases of the spin waves are not synchronized[26,28], reducing the efficiency of mutual coherence stimulation between the two channels, which limits the bandwidth of the NHPA process to ~100 Hz. To overcome this limitation, a periodic drive is used to spectrally tailor our system with synthetic levels, through the photon-assisted Floquet coherent transition. We note that Floquet engineering has been recently employed in trapped ions and atomic gases to realize discrete time crystals[39], topological band structures[40], and Floquet masers[41]. However, to the best of our knowledge, it has not been used for controlling quantum correlations in any non-Hermitian system.

Next, we show that periodic modulations help expand the single-color quantum nonreciprocity to multi-color sidebands, thereby increasing the bandwidth of our system. For this purpose, we add an oscillating magnetic field $B_1 \cos(\omega_1 t)$ to the static field $B_0$ along z-axis to realize periodically driven Zeeman levels. According to Floquet theorem[42], the evolution dynamics of a periodically driven two-level system can be described with Floquet quasi-energy states $|\pm\rangle_n = \sum_m \mathcal{J}_{n-m}(\pm \pi \gamma B_1/\omega_1)|\pm, m\rangle$ and the associated ladder-level energies[40]: $E_{\pm,n} = \pm\omega_0 + n\omega_1$, where $\gamma$ is the gyromagnetic ratio of $^{87}$Rb. Here, $\mathcal{J}_{n-m}$ is the Bessel function of the first kind of order $(n-m)$, and $|\pm, m\rangle$ indicates that the spin is in the up $|+\rangle$ or down $|-\rangle$ state with photon number $m$ in the periodic driving field (Fig. 3**a**). When the system is configured as NHPA and only a control beam is used in CH2, the periodically driven EIT spectra of CH2 exhibit first- and second-order sidebands (Fig. 3**b**), implying multimode oscillations at tunable frequencies of transition between synthetic quasi-energy states. Here, both control and probe beams are switched on in CH1, and CH1's probe frequency is swept around the two-photon resonance location of the carrier ($\Delta = 0$). The non-zero frequency of the carrier EIT center is attributed to the AC Stark shift from the control beam. We calibrate the peak amplitude of the carrier and the prominent first-order sidebands by fitting with a zero- and first-order Bessel function of the first kind $\mathcal{J}_{0(1)}(k_U/\nu_1)$ as function of the modulation frequency $\nu_1$ with $\omega_1 = 2\pi\nu_1$, confirming the Floquet theorem in our system. Here, $k_U$ is a parameter proportional to the modulation depth $U$. We observe a multimode feature with carrier and first-order sidebands in the EIT spectra for $\nu_1 = 3$ kHz and $U = 6$ V. To explore quantum correlations in this regime, we perform quantum noise measurements via joint polarization homodyne detection. Intriguingly, we find that the quantum nonreciprocity indeed emerges at different frequencies by tuning the modulation depth (Fig. 3**c**). Specifically, nonreciprocal quantum correlations occur at 299 kHz and $299 \pm 3$ kHz for $U = 5.5$ V and $\nu_1 = 3$ kHz. Moreover, for the case of $U = 7.6$ V and $\nu_1 = 2$ kHz, we also observe the multicolor quantum nonreciprocity with the first, second, and third sidebands, in which the carrier locates at 298.3 kHz.

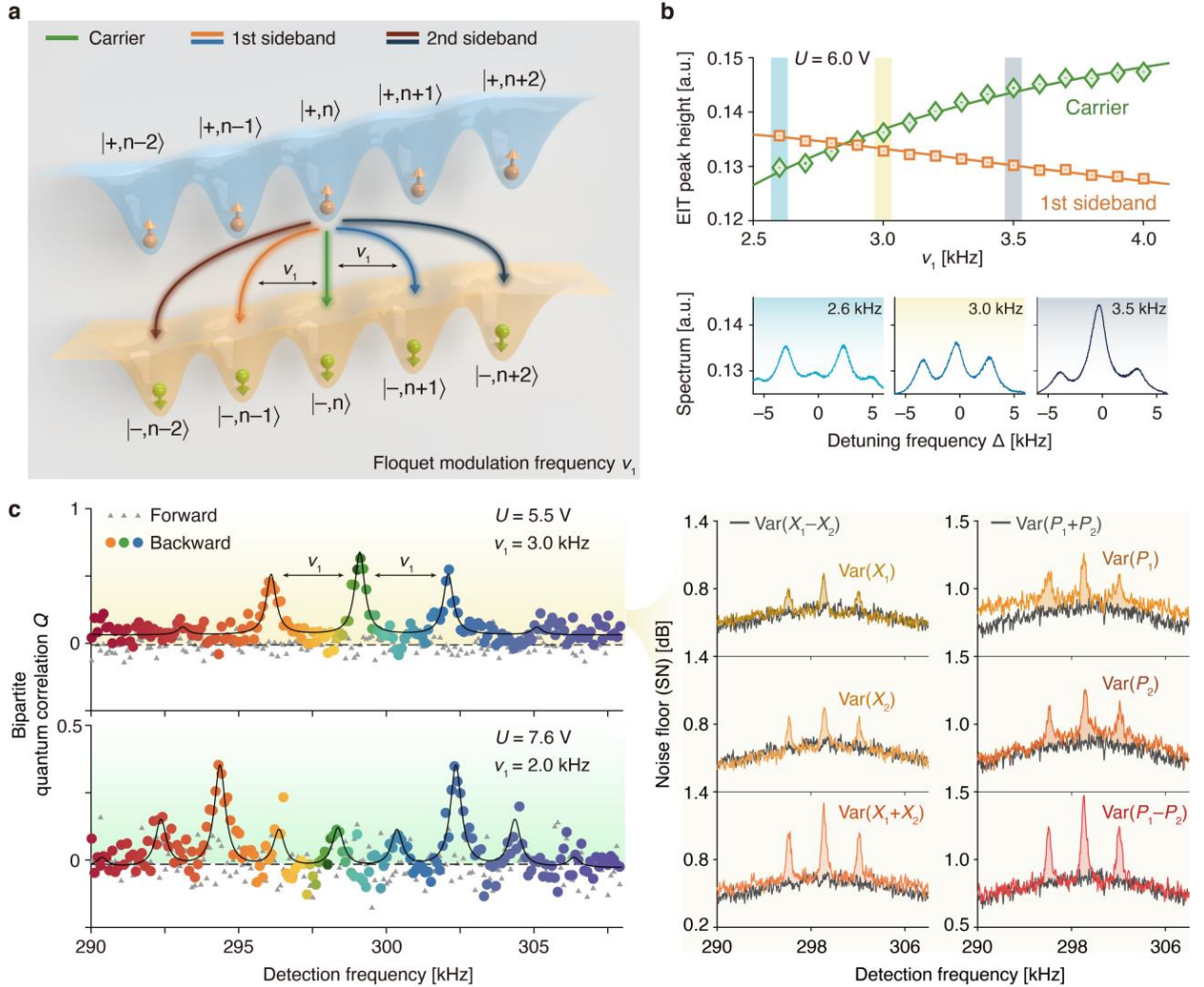

**Fig. 3 | Multicolor quantum nonreciprocity induced by Floquet engineering. a**, The states modulated by a periodic drive show the two-photon transitions $|+, n\rangle \to |-, n\rangle$, $|+, n\rangle \to |-, n \pm 1\rangle$, and $|+, n\rangle \to |-, n \pm 2\rangle$, which indicate the carrier, first- and second-sidebands in the EIT spectra, respectively. **b**, Peak amplitude of the carrier (green) and first-order sidebands (orange) as a function of the modulation frequency $v_1$ with the modulation depth kept fixed ($U = 6.0$ V). Green and orange curves illustrate the theoretical fitting with Bessel function. Here, $\Delta$ is the two-photon detuning with respect to the carrier. **c**, Multicolor quantum nonreciprocity, compared to the monochromatic one shown in Fig. 2c, occurs by tuning $U$ and $v_1$. The colored markers are experimentally measured results, while the solid (dashed) black curve is theoretical fitting for backward (forward) case. The noise spectra for $U = 5.5$ V and $v_1 = 3$ kHz are plotted in the right panels.

In conclusion, we have demonstrated a chirality-induced quantum effect, i.e., nonreciprocal optical two-channel correlations by using a non-Hermitian macroscopic system. We find that by only flipping the input direction of the same laser through the same atomic vapor, quantum two-channel correlations can be switched on and off. Also, we demonstrate an efficient way to achieve multi-color quantum nonreciprocity by using Floquet engineering, offering opportunities to reveal new features or functionality of non-Hermitian chiral systems with synthetic dimensions. Our findings establish a unique link of two important asymmetries, chirality and nonreciprocity, in the quantum cross-correlation level, highlight the roles of dissipative interactions and vacuum fields in observing such a unique link, and present a step towards exploring multipartite quantum effects with chiral non-Hermitian systems, for a wide range of applications in e.g. chiral quantum engineering[43], one-way quantum information processing or in-memory computing[44], as well as nonreciprocal quantum sensing[45] or imaging[46].

We note that classical nonreciprocity, i.e., one-way transmission of classical light, plays a key role for improving the performance of optical networks by preventing signal interference, and by enabling essential devices used in optical signal processing, such as optical isolators and circulators[8,47,48]. Thus, quantum nonreciprocity is also expected to have important applications in e.g., achieving one-way quantum network (without quantum information backflow)[5], making nonreciprocal quantum sensors[45], or protecting quantum states against random noises[23]. Moreover, we note that classical nonreciprocity was already utilized as a powerful resource for sensing, which allows one to exceed the fundamental bounds constraining any conventional, reciprocal sensor[49]. We expect that quantum nonreciprocity can also be used to improve the performance of quantum sensors by enhancing one-way quantum correlations against backscattering losses[45]. We also believe that the fundamental link between chirality and quantum nonreciprocity, as revealed here, can provide a deeper understanding of both concepts, and may stimulate future efforts for building and using devices and systems that exploit quantum directional effects.

**References**

1. Brown, J. M. & Davies, S. G. Chemical asymmetric synthesis. *Nature* **342**, 631-636 (1989).

2. Wang, Y., Xu, J., Wang, Y. & Chen, H. Emerging chirality in nanoscience. *Chem. Soc. Rev.* **42**, 2930-2962 (2013).

3. Behera, P. et al. Electric field control of chirality. *Sci. Adv.* **8**, eabj8030 (2022).

4. Zhang, X., Liu, Y., Han, J., Kivshar, Y. & Song, Q. Chiral emission from resonant metasurfaces. *Science* **377**, 1215-1218 (2022).

5. Lodahl, P. et al. Chiral quantum optics. *Nature* **541**, 473-480 (2017).

6. Bliokh, K. Y., Smirnova, D. & Nori, F. Quantum spin Hall effect of light. *Science* **348**, 1448-1451 (2015).

7. Bliokh, K. Y., Leykam, D., Lein, M. & Nori, F. Topological non-Hermitian origin of surface Maxwell waves. *Nat. Commun.* **10**, 580 (2019).

8. Sounas, D. L. & Alù, A. Non-reciprocal photonics based on time modulation. *Nat. Photon.* **11**, 774-783 (2017).

9. Junge, C., O' Shea, D., Volz, J. & Rauschenbeutel, A. Strong coupling between single atoms and nontransversal photons. *Phys. Rev. Lett.* **110**, 213604 (2013).

10. Mitsch, R., Sayrin, C., Albrecht, B., Schneeweiss, P. & Rauschenbeutel, A. Quantum state-controlled directional spontaneous emission of photons into a nanophotonic waveguide. *Nat. Commun.* **5**, 5713 (2014).

11. Luxmoore, I. J. et al. Interfacing spins in an InGaAs quantum dot to a semiconductor waveguide circuit using emitted photons. *Phys. Rev. Lett.* **110**, 037402 (2013)

12. Sapienza, L. et al. Cavity quantum electrodynamics with Anderson localized modes. *Science* **327**, 1352-1355 (2010).

13. Volz, J., Scheucher, M., Junge, C. & Rauschenbeutel, A. Nonlinear $\pi$ phase shift for single fibre-guided photons interacting with a single resonator-enhanced atom. *Nat. Photon.* **8**, 965-970 (2014).

14. Rosenblum, S. et al. Extraction of a single photon from an optical pulse. *Nat. Photon.* **10**, 19-22 (2016).

15. Douglas, J. S. et al. Quantum many-body models with cold atoms coupled to photonic crystals. *Nat. Photon.* **9**, 326-331 (2015).

16. Hung, C. L., Gonzalez-Tudela, A., Cirac, J. I. & Kimble, H. J. Quantum spin dynamics with pairwise-tunable, long-range interactions. *Proc. Natl Acad. Sci. USA* **113**, E4946-E4955 (2016).

17. Shomroni, I. et al. All-optical routing of single photons by a one-atom switch controlled by a single photon. *Science* **345**, 903-906 (2014).

18. Scheucher, M., Hilico, A., Will, E., Volz, J. & Rauschenbeutel, A. Quantum optical circulator controlled by a single chirally coupled atom. *Science* **354**, 1577-1580 (2016).

19. Söllner, I. et al. Deterministic photon–emitter coupling in chiral photonic circuits. *Nat. Nanotechnol.* **10**, 775-778 (2015).

20. Dong, M.-X. et al. All-optical reversible single-photon isolation at room temperature. *Sci. Adv.* **7**, eabe8924 (2021).

21. Dötsch, H. et al. Applications of magneto-optical waveguides in integrated optics: review. *J. Opt. Soc. Am. B* **22**, 240-253 (2005).

22. Huang, R., Miranowicz, A., Liao, J., Nori, F. & Jing, H. Nonreciprocal photon blockade. *Phys. Rev. Lett.* **121**, 153601 (2018).

23. Jiao, Y.-F. et al. Nonreciprocal Optomechanical Entanglement against Backscattering Losses. *Phys. Rev. Lett.* **125**, 143605 (2020).

24. Graf, A. et al. Nonreciprocity in Photon Pair Correlations of Classically Reciprocal Systems. *Phys. Rev. Lett.* **128**, 213605 (2022).

25. del Pino, J., Slim, J. J. & Verhagen, E. Non-Hermitian chiral phononics through optomechanically induced squeezing. *Nature* **606**, 82-87 (2022).

26. Peng, P. et al. Anti-parity-time symmetry with flying atoms. *Nat. Phys.* **12**, 1139-1145 (2016).

27. Wen R. et al. Non-Hermitian Magnon-Photon Interference in an Atomic Ensemble. *Phys. Rev. Lett.* **122**, 253602 (2019).

28. Cao, W. et al. Reservoir-Mediated Quantum Correlations in Non-Hermitian Optical System. *Phys. Rev. Lett.* **124**, 030401 (2020).

29. Lu, X., Cao, W., Yi, W., Shen, H. & Xiao, Y. Nonreciprocity and Quantum Correlations of Light Transport in Hot Atoms vis Reservoir Engineering. *Phys. Rev. Lett.* **126**, 223603 (2021).

30. Zhang, S. et al. Thermal-motion-induced non-reciprocal quantum optical system. *Nat. Photon.* **12**, 744-748 (2018).

31. Liang, C. et al. Collision-Induced Broadband Optical Nonreciprocity. *Phys. Rev. Lett.* **125**, 123901 (2020).

32. Verstraete, F., Wolf, M. M. & Cirac, J. I. Quantum computation and quantum-state engineering driven by dissipation. *Nat. Phys.* **5**, 633-636 (2009).

33. Schindler, P. et al. Quantum simulation of dynamical maps with trapped ions. *Nat. Phys.* **9**, 361-367 (2013).

34. Yang, W. et al. A self-biased non-reciprocal magnetic metasurface for bidirectional phase modulation. *Nat. Electron.* **6**, 225-234 (2023).

35. Liu, M. et al. Broadband mid-infrared non-reciprocal absorption using magnetized gradient epsilon-near-zero thin films. *Nat. Mater.* **22**, 1196-1202 (2023).

36. Wang, J. et al. Light-induced fictitious magnetic fields for quantum storage in cold atomic ensembles. *Phys. Rev. Research* **6**, L042002 (2024).

37. Adesso, G. & Datta, A. Quantum versus Classical Correlations in Gaussian States. *Phys. Rev. Lett.* **105**, 030501 (2010).

38. Giorda, P. & Paris, M. G. A. Gaussian Quantum Discord. *Phys. Rev. Lett.* **105**, 020503 (2010).

39. Zhang, J. et al. Observation of a discrete time crystal. *Nature* **543**, 217-220 (2017).

40. Weitenberg, C. & Simonet, J. Tailoring quantum gases by Floquet engineering. *Nat. Phys.* **17**, 1342-1348 (2021).

41. Jiang, M., Su, H., Wu, Z., Peng, X. & Budker, D. Floquet maser. *Sci. Adv.* **7**, eabe0719 (2021).

42. Grifoni, M. & Hänggi, P. Driven quantum tunneling. *Phys. Rep.* **304**, 229-354 (1998).

43. Zhang, Y. et al. Chirality logic gates. *Sci. Adv.* **8**, eabq8246 (2022).

44. Pintus, P. et al. Integrated non-reciprocal magneto-optics with ultra-high endurance for photonic in-memory computing. *Nat. Photon.* (2024).

45. Wang, J., Zhang, Q. & Jing, H. Quantum advantage of one-way squeezing in weak-force sensing. *Appl. Phys. Rev.* **11**, 031409 (2024).

46. Vernière, C. & Defienne, H. Hiding Images in Quantum Correlations. *Phys. Rev. Lett.* **133**, 093601 (2024).

47. Nagulu, A., Reiskarimian, N. & Krishnaswamy, H. Non-reciprocal electronics based on temporal modulation. *Nat. Electron.* **3**, 241–250 (2020).

48. Fruchart, M. et al. Non-reciprocal phase transitions. *Nature* **592**, 363–369 (2021).

49. Lau, H.-K. & Clerk, A. A. Fundamental limits and non-reciprocal approaches in non-Hermitian quantum sensing. *Nat. Commun.* **9**, 4320 (2018).

# Methods

**Experiment setup**

We use a paraffin-coated cylindrical cell, with diameter of 2.5 cm and length of 7.5 cm, containing isotopically enriched $^{87}$Rb vapor. The cell is mounted inside a magnetic shielding, in which a set of coils provides precise control over the internal longitudinal magnetic field. The number density of Rb atoms is controlled by adjusting the temperature of the pull-off of the cell (54 °C in this work). A diode laser is tuned to the D1 line of $^{87}$Rb and is split into control and probe beams. As shown in Extended Data Fig. 1, in each channel, a relatively strong control beam completely overlaps with the weak probe beam (or vacuum mode) with orthogonal circular polarizations (5 mm in diameter), forming the $\Lambda$-type electromagnetically induced transparency (EIT) configuration with Zeeman sublevels. In practice, we always stabilize the pump laser frequency to cross peak of $|F = 2\rangle$ to $|F' = 1\rangle$ and $|F = 2\rangle$ to $|F' = 2\rangle$ of D1 transition via the locking technique based on saturation absorption spectroscopy. Intrinsic ballistic motion of thermal atoms plays a key role for realizing the dissipative coupling between two optical channels: the atoms can move back and forth between the channels many times within coherence lifetime given the high atomic velocity (160 m/s) and the small diameter of the atomic cell (only 2.5 cm). This is enabled by the anti-relaxation coating inside the vapor cell, which allows atoms to undergo hundreds or thousands of wall collisions with little demolition of their internal quantum state.

In quantum nonreciprocal experiment, we switch off the input probe fields in both channels, replacing them with coherent vacuum. In the forward scattering process, the control and the vacuum state in the optical mode satisfy the energy and momentum conservation condition. Specifically, when the control beam resonantly couples to the transition $|1\rangle \to |3\rangle$ ($|2\rangle \to |3\rangle$), the frequency of this optical mode fulfills the two-photon resonance condition in the $\Lambda$-type three-level scheme, i.e., the optical mode resonantly couples to $|2\rangle \to |3\rangle$ ($|1\rangle \to |3\rangle$). In addition, the polarization of this optical mode is orthogonal to that of the control beam. Due to the forward scattering process, the newly generated quantum light field in this particular optical mode corresponding to the transition $|3\rangle \to |2\rangle$ ($|3\rangle \to |1\rangle$) propagates along the direction of the control field in each channel.

To extract quantum noise and correlation information, each of the output beams of the cell is detected by a separate polarization homodyne measurement setup, which is composed of a polarization beam-splitter and home-built balanced homodyne detector. The noise power of the amplified difference photocurrents in the balanced homodyne detector is recorded with a spectrum analyzer. The circularly polarized control fields also play the role of the local oscillators of the homodyne detectors at the output for quantum noise measurements of the probe fields. The optical transmission between the end window of the vapor cell and the balanced photodetector is 92%. The quantum efficiency of the balanced photodetector is 92%.

Bipartite quantum correlation or quantum correlation between two channels are measured by a joint homodyne detection, typically used in the measurement of continuous variable entanglement[28], consisting of two sets of polarization homodyne detectors and two radio-frequency power splitters/combiners. More details can be found in Refs. [28,50].

In classical nonreciprocal experiment, to obtain the spectra of EIT carried by the probe, circularly polarized probe and control beams after the cell are converted to orthogonal linear polarization by quarter-wave plates, and directed to the polarization beam-splitter, as shown in the inset of Extended Data Fig. 1. The frequency of the probe in CH1 is swept by scanning the driving frequency on the second acousto-optic modulator in Extended Data Fig. 1. Here, the +1-order diffracted beam of the first acousto-optic modulator (at 80 MHz) is fed into the second acousto-optic modulator, and the −1-order diffraction of the second acousto-optic modulator (at 80 MHz+Δ) is utilized as CH1's probe, to ensure the accuracy of the frequency detuning with respect to the control.

There are three main differences between classical and quantum nonreciprocal effects: (i) The physical principles are different. The mechanism behind the classical nonreciprocity is the Doppler shift induced by the moving atoms, which is independent of the interaction between two channels. However, the principle behind the quantum nonreciprocity is the chirality of the light beam interacting with the atoms. (ii) The experimental systems are different. An additional probe light is required in the experiments of classical nonreciprocal effects for the measurements of transmission spectra. For measurements of quantum noise and correlations, we remove the input probes and let them be the vacuum in the experiments of quantum nonreciprocity. (iii) The measured observables are different. In the classical nonreciprocal effects, we measure the transmission spectra of the probe. For the experiments of quantum nonreciprocity, we measure the quantum fluctuations in the output vacuum modes, as well as the quantum correlations between the two channels.

**Chirality of light**

Consider the plane-wave solution to Maxwell's equations given by:
$$\vec{E}(\vec{r},t) = \vec{E}_0 \exp[i(\vec{k}\cdot\vec{r} - \omega t)], \tag{1}$$

The wave vector $\vec{k}$ specifies the direction of propagation, and $\vec{k}$ and $\vec{E}_0$ are perpendicular. For $\vec{k}$ in $+z$ direction, we can write two orthogonal complex electric field components in the $x$ and $y$ directions respectively:
$$\vec{E}_x(z,t) = E_0 \cos(\omega t - kz), \quad \vec{E}_y(z,t) = E_0 \sin(\omega t - kz). \tag{2}$$

For left-hand circularly polarized (LHCP) light, the amplitudes of these two components are equal, and the phase difference is $\pi/2$. When light propagates along $+z$ direction, these two components can be merged into a single complex electric field: $\vec{E}_0 = (E_0, E_0 e^{i\frac{\pi}{2}}, 0) = (E_0, iE_0, 0)$,
$$\vec{E} = (E_0 e^{i(kz-\omega t)}, iE_0 e^{i(kz-wt)}, 0). \tag{3}$$

Taking the real part gives the actual electric field:
$$\mathrm{Re}[\vec{E}]_{-} = (E_0 \cos(kz-\omega t), -E_0 \sin(kz-\omega t), 0). \tag{4}$$

where "−" indicates LHCP light. By ignoring the time-dependent term, the LHCP light propagating in $+z$ direction is written as:
$$\mathrm{Re}[\vec{E}]_{-} = (E_0 \cos(kz), -E_0 \sin(kz), 0). \tag{5}$$

Similarly, the right-hand circularly polarized (RHCP) light propagating in $+z$ direction is given by:
$$\mathrm{Re}[\vec{E}]_{+} = (E_0 \cos(kz-\omega t), E_0 \sin(kz-\omega t), 0). \tag{6}$$

If we keep using $\vec{k}$ as the wave vector, when the RHCP light propagats in $-z$ direction, it can be described as:

$$\text{Re}[\vec{E}]_+' = (E_0 \cos(-kz - \omega t), E_0 \sin(-kz - \omega t), 0). \tag{7}$$

By ignoring the time-dependent term, the RHCP light propagating in $-z$ direction is given by:

$$\begin{aligned}\text{Re}[\vec{E}]_+' &= (E_0 \cos(-kz), E_0 \sin(-kz), 0), \\ &= (E_0 \cos(kz), -E_0 \sin(kz), 0),\end{aligned} \tag{8}$$

which is same as the LHCP light propagating in $+z$ direction. Therefore, two controls with the same right-hand circular polarizations propagating in the forward case, the 'atoms' see the same chirality of the control beams. However, in the backward case, the 'atoms' see the opposite chirality of the control beams, since the reversed right-hand circularly polarized light is similar to the left-hand circularly polarized light in the forward case.

**Measuring the quantum noise for the quadratures of light**

The quantum state of light can be characterized by the Stokes operators $\hat{S}_x$, $\hat{S}_y$, and $\hat{S}_z$, which are given by the differences of the number operators $\hat{n}_{\text{polarization}}$ of photons polarized in different orthogonal bases. In the circular polarization basis of $\hat{a}_R$ (right-hand circular polarization) and $\hat{a}_L$ (left-hand circular polarization), we have:

$$\hat{S}_x = -\tfrac{1}{2}(\hat{a}_R^\dagger \hat{a}_L + \hat{a}_L^\dagger \hat{a}_R), \tag{9}$$

$$\hat{S}_y = -\tfrac{1}{2i}(\hat{a}_R^\dagger \hat{a}_L - \hat{a}_L^\dagger \hat{a}_R), \tag{10}$$

$$\hat{S}_z = -\tfrac{1}{2}(\hat{a}_R^\dagger \hat{a}_R - \hat{a}_L^\dagger \hat{a}_L), \tag{11}$$

In the experiments, we use $\sigma_\pm$ circularly polarized coherent light to interact with the atomic ensembles (input probes are vacuum), which means $\hat{S}_z$ can be treated as a large classical value proportional to the photon flux $\Phi = P/\hbar\omega$. Here, $P$ represents the optical power, and $\hbar\omega$ is the energy of a single photon. The quantum variables $\hat{S}_x$ and $\hat{S}_y$ are the physical variables we are interested in, and they have zero mean value[51]. The control light of CH1 is right-hand circularly polarized (RHCP) and that of CH2 is left-hand circularly polarized (LHCP). Thus, in CH1 we have

$$\hat{X}_1 = -\hat{S}_x/\sqrt{S_z} = \tfrac{1}{\sqrt{2}}(\hat{a}_L + \hat{a}_L^\dagger), \quad \hat{P}_1 = -\hat{S}_y/\sqrt{S_z} = \tfrac{1}{i\sqrt{2}}(\hat{a}_L - \hat{a}_L^\dagger), \tag{12}$$

and in CH2, we have

$$\hat{X}_2 = -\hat{S}_x/\sqrt{S_z} = \tfrac{1}{\sqrt{2}}(\hat{a}_R + \hat{a}_R^\dagger), \quad \hat{P}_2 = -\hat{S}_y/\sqrt{S_z} = \tfrac{1}{i\sqrt{2}}(\hat{a}_R - \hat{a}_R^\dagger). \tag{13}$$

The sign difference in $\hat{P}_1$ and $\hat{P}_2$ has taken into account the fact that $\hat{S}_z$ in the two channels have opposite signs, and $[\hat{X}_1, \hat{P}_1] = [\hat{X}_2, \hat{P}_2] = i$ ($\hbar = 1$) is satisfied.

At the output of the two channels, we use the combination of $\lambda/2$ waveplate and $\lambda/4$ waveplate to rotate the Stokes vectors $\hat{S}_x$, $\hat{S}_y$, and $\hat{S}_z$, so that the desired quadrature of light $\hat{X}$ or $\hat{P}$ is detected by the balanced detector. The noise power spectra of $X_1(\omega)$, $X_2(\omega)$, $P_1(\omega)$, and $P_2(\omega)$ are then obtained from the spectrum analyzer.

**Extracting the Gaussian discord**

Gaussian quantum correlations beyond entanglement are captured by the measure of Gaussian discord. In a bipartite system, the total amount of correlations (classical and quantum) is given by the von Neumann mutual information

$$I(\rho_{AB}) = S(\rho_A) + S(\rho_B) - S(\rho_{AB}), \tag{14}$$

where $S(\rho)$ is the von Neumann entropy, and $\rho_{A(B)}$ is the reduced density matrix of the subsystem $A$ ($B$). Another measure of mutual information is

$$J_A(\rho_{AB}) = S(\rho_A) - \inf_{\sigma_M} S(\rho_A|\sigma_M), \tag{15}$$

which quantifies only the amount of classical correlations and can be extracted by a Gaussian measurement. Here, $\sigma_M$ is the covariance matrix of the measurement on mode $B$. As it only indicates the classical correlations, the difference of above two definitions of mutual information is a measure of Gaussian quantum correlation that is referred to as the Gaussian quantum discord[37,38], i.e.,

$$\mathfrak{D}_A = I(\rho_{AB}) - J_A(\rho_{AB}). \tag{16}$$

The discord of a bipartite system can be calculated from its covariance matrix, which can be reconstructed using single/joint homodyne detection. The $4 \times 4$ covariance matrix for the state $\rho_{AB}$ written in the standard form is

$$\sigma_{AB} = \begin{pmatrix} \alpha & \gamma \\ \gamma^T & \beta \end{pmatrix}, \tag{17}$$

where the submatrices $\alpha, \beta$, and $\gamma$ are defined as

$$\alpha = \text{diag}[\text{Var}(X_A), \text{Var}(P_A)], \tag{18}$$
$$\beta = \text{diag}[\text{Var}(X_B), \text{Var}(P_B)], \tag{19}$$
$$\gamma = \text{diag}[\text{Cov}(X_A, X_B), \text{Cov}(P_A, P_B)], \tag{20}$$

with

$$\text{Cov}(\hat{O}_1, \hat{O}_2) = \tfrac{1}{2}\langle \hat{O}_1 \hat{O}_2 + \hat{O}_2 \hat{O}_1 \rangle, \tag{21}$$

The covariance value can be obtained through joint homodyne detection using power splitters or combiners as

$$\text{Cov}(X_A, X_B) = \tfrac{1}{2}[\text{Var}(X_A + X_B) - \text{Var}(X_A) - \text{Var}(X_B)], \tag{22}$$
$$\text{Cov}(P_A, P_B) = -\tfrac{1}{2}[\text{Var}(P_A - P_B) - \text{Var}(P_A) - \text{Var}(P_B)]. \tag{23}$$

With the covariance matrix $\sigma_{AB}$ in hand, one can calculate the discord using

$$\mathfrak{D}(\sigma_{AB}) = h(\sqrt{I_2}) - h(\nu_-) - h(\nu_+) + h(\sqrt{E^{\min}}), \tag{24}$$

where

$$h(x) = \tfrac{1}{2}(x+1)\log\left[\tfrac{1}{2}(x+1)\right] - \tfrac{1}{2}(x-1)\log\left[\tfrac{1}{2}(x-1)\right], \tag{25}$$

$$\nu_\pm^2 = \tfrac{1}{2}\left[\delta \pm \sqrt{\delta^2 - 4I_4}\right], \quad \delta = I_1 + I_2 + 2I_3, \tag{26}$$

$$I_1 = \det[\alpha], \quad I_2 = \det[\beta], \quad I_3 = \det[\gamma], \quad I_4 = \det[\sigma_{AB}]. \tag{27}$$

If the following condition is satisfied

$$(I_4 - I_1 I_2)^2 \leq I_3^2 (I_2 + 1)(I_4 + 1), \tag{28}$$

we have

$$E^{\min} = \left[2I_3^2 + (I_2 - 1)(I_4 - 1) + 2|I_3|\sqrt{I_3^2 + (I_2 - 1)(I_4 - 1)}\right]/(I_2 - 1)^2; \tag{29}$$

otherwise, we have

$$E^{\min} = \left[I_1 I_2 - I_3^2 + I_4 - \sqrt{I_3^4 + (I_4 - I_1 I_2)^2 - 2I_3^2(I_4 + I_1 I_2)}\right]/2I_2. \tag{30}$$

The above derivations provide the results of Gaussian quantum discord calculated in Fig. 2 of the main text. In addition, we provide the value of Gaussian quantum discord with error bar in Fig. S2 of Supplementary Information. The error bar of such a small number, which ensures the nonzero quantity of quantum discord (see more discussions about Gaussian quantum discord in Sec. S.3 of Supplementary Information).

**Floquet engineering**

In addition to the common magnetic field produced by the coils, an oscillating magnetic field at the frequency of $\nu_1$ is applied in the same direction to modulate the Zeeman levels. A time-periodic modulation forces a quantum state in each spatial channel dressed by all harmonics of the driving frequency, and the different harmonics- which manifests as the sidebands ($n = 0, 1, 2, …$) of the atomic spin waves, as shown in Fig. 3 of the main text. More interestingly, the multicolor quantum nonreciprocity, i.e., nonreciprocal quantum correlations occurring at different frequencies, are observed by using this periodic modulation.

Furthermore, such a method could enable the creation and manipulation of a dissipative coupling between two arbitrary Floquet atomic sidebands emerged in two spatially separated optical channels. By tuning the frequency difference between control beams in two channels to satisfy $\Delta_0 \approx \pm n\nu_1$ with the frequency difference of atomic spin waves $\Delta_0$, one can realize the dissipative coupling between collective spin waves associated with sideband index $n_1$ and $n_2$ ($n = n_1 - n_2$), respectively in CH1 and CH2. Along this way, we demonstrated the propagation-direction-dependent chiral interactions between different sidebands with $n = 1$, as shown in Extended Data Fig. 5.

**References**

50. Davidovich, L. Sub-Poissonian processes in quantum optics. *Rev. Mod. Phys.* **68**, 127-173 (1996).
51. Shen, H. "Spin squeezing and entanglement with room temperature atoms for quantum sensing and communication," thesis, University of Copenhagen (2015).
**Data availability**

All data is available in the main text or the supplementary materials.

**Acknowledgments**

The authors are grateful to Jing Zhang, Min Jiang and Ying Hu for fruitful discussions. This work is supported by National Key Research Program of China under Grant No. 2020YFA0309400, and NNSFC under Grants No. 12222409 and 12174081, 11974228, and the Key Research and Development Program of Shanxi Province under Grant No. 202101150101025. H. Shen

acknowledges the financial support from the Royal Society Newton International Fellowship Alumni follow-on funding (AL201024) of the United Kingdom. H. Jing is supported by the NSFC (Grant No. 11935006, 12421005), the Sci-Tech Innovation Program of Hunan Province (Grant No. 2020RC4047), the National Key R&D Program (Grant No. 2024YFE0102400), and the Hunan Major Sci-Tech Program (Grant No. 2023ZJ1010). R. Huang is supported by the RIKEN Special Postdoctoral Researchers (SPDR) program. Ş.K. Özdemir acknowledges the Air Force Office of Scientific Research (AFOSR) Multidisciplinary University Research Initiative (MURI) Award on Programmable systems with non-Hermitian quantum dynamics (Award No. FA9550-21-1-0202). F. Nori is supported in part by: Nippon Telegraph and Telephone Corporation (NTT) Research, the Japan Science and Technology Agency (JST) [via the CREST Quantum Frontiers program Grant No. JPMJCR24I2, the Quantum Leap Flagship Program (Q-LEAP), and the Moonshot R&D Grant No. JPMJMS2061], and the Office of Naval Research (ONR) Global (via Grant No. N62909-23-1-2074).

**Author contributions**

B.C., H.B., H.J. and H.S. conceived the idea. Z.Z., Z.X., X.L., F.Z. and D.L. conducted the experiment and analyzed the data together with all other authors. Z.Z., Z.X., R.H., Y.X., Ş.K.Ö., F.N., H.J. and H.S. wrote the manuscript with contributions from all the other authors.

**Competing interests**

Authors declare no competing interests.

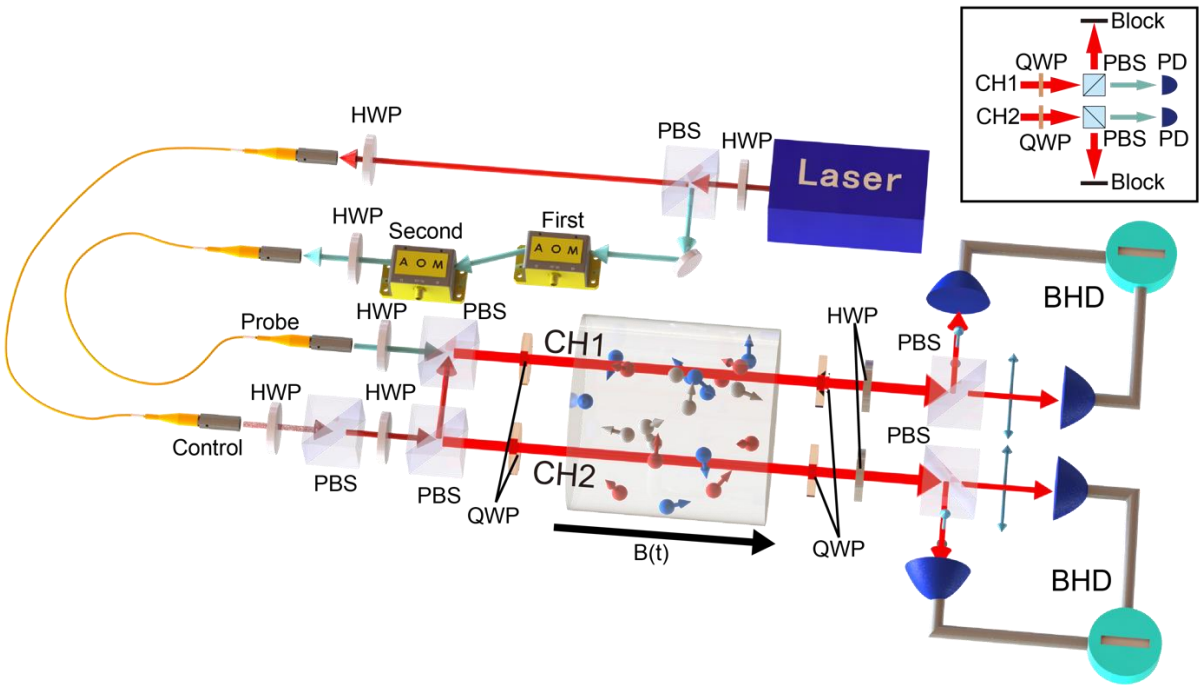

**Extended Data Fig. 1 | Experimental setup.** AOM, acousto-optic modulator; PBS, polarization beam-splitter; BHD, balanced homodyne detector; −, subtractor; HWP, half-wave plate; QWP, quarter-wave plate; PD, photodetector.

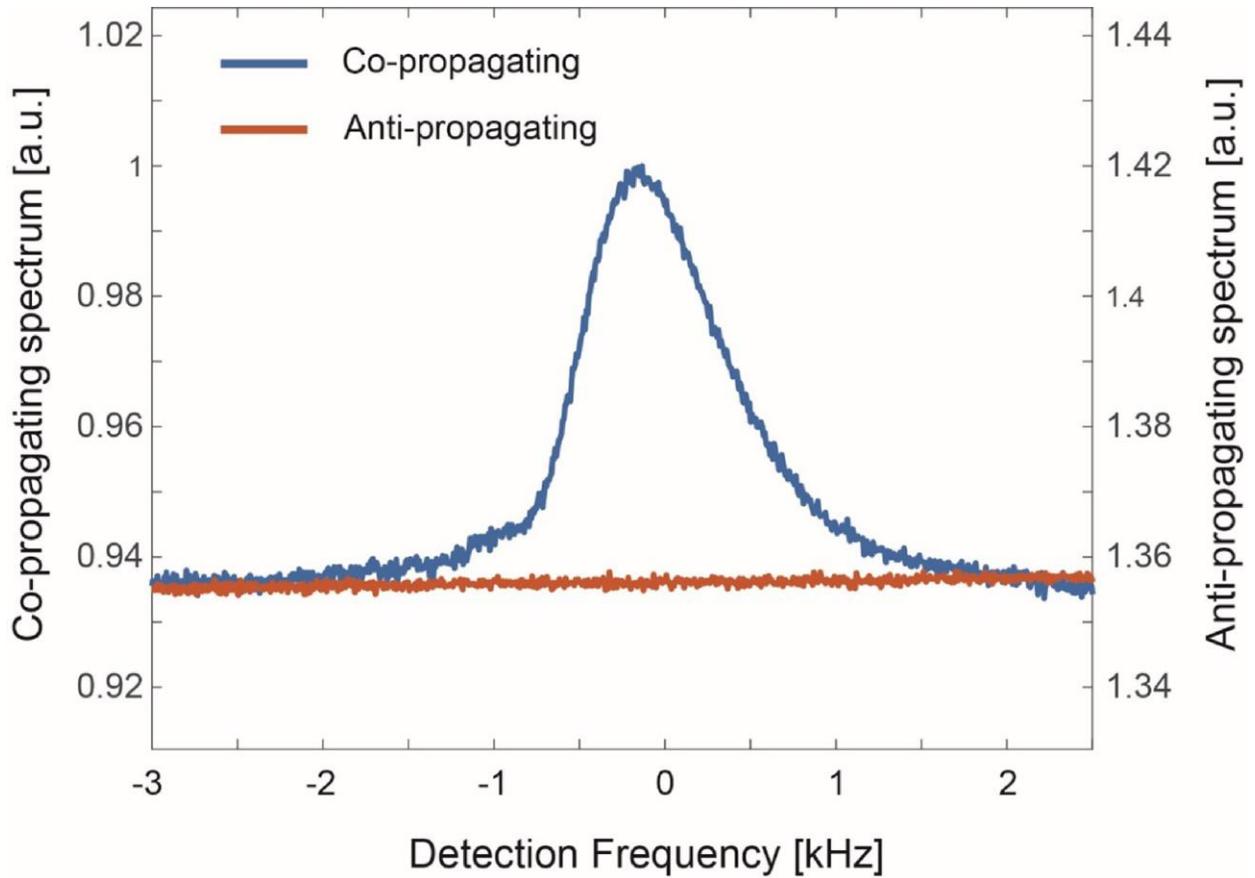

**Extended Data Fig. 2 | The measured data of classical nonreciprocity introduced by Doppler shift.** The experimental schematic and three-level Λ-type EIT configurations are shown in Fig. 1. When the probe in CH2 propagates in the forward direction, there is a classical EIT spectrum in CH2. When it propagates in the backward direction, no EIT spectrum exists.

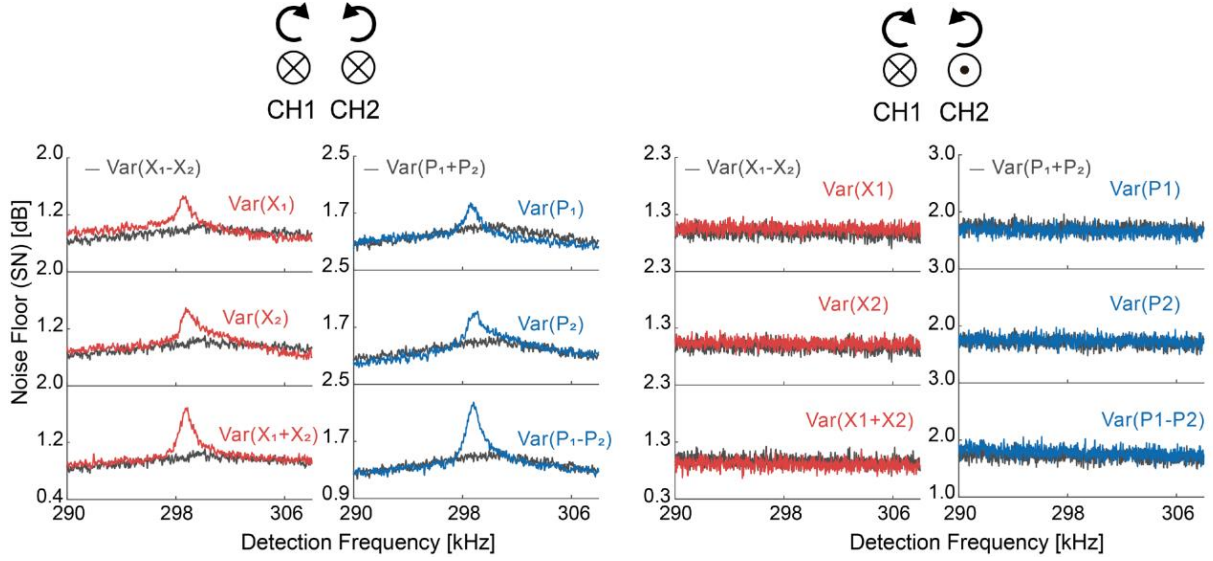

**Extended Data Fig. 3 | Nonreciprocal bipartite quantum correlations under the reversed polarizations configuration.** When the polarizations of the fields are orthogonal, nonlinear non-Hermitian parametric-amplifier (NHPA) and linear dissipative beam-splitter (DBS) interactions are created for co-propagating (left panels) and counter-propagating (right panels) fields, respectively. Here, ↻ and ↺ represent RHCP and LHCP polarizations, respectively. The markers of ⊗ and ⊙ denote the propagation direction. Quantum correlation induced by NHPA is clearly seen in the measured quantum noises, while no quantum correlation can be observed with DBS. This direction-dependent quantum correlation (i.e., quantum nonreciprocity) is further confirmed by the Gaussian quantum discord with $\mathfrak{D}_1 = 0.9 \times 10^{-3}$.

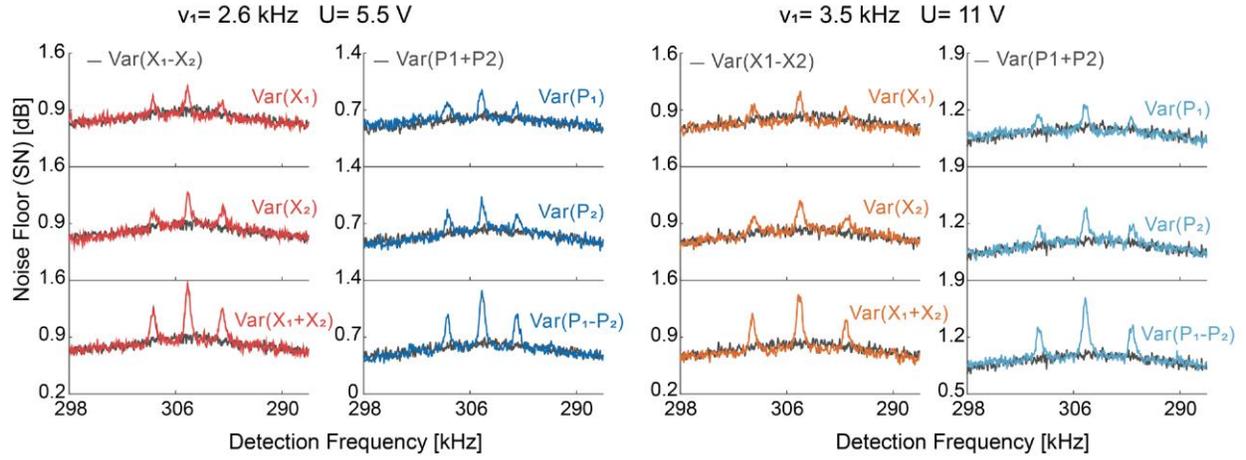

**Extended Data Fig. 4 | Multicolor bipartite quantum correlations with different periodic modulations.** Quantum correlations with broader bandwidth can be obtained by using periodic modulation (i.e., Floquet techniques), where the carrier and first-sideband are shown in the noise spectra with different modulation frequencies $\nu_1$ and modulation depths $U$.

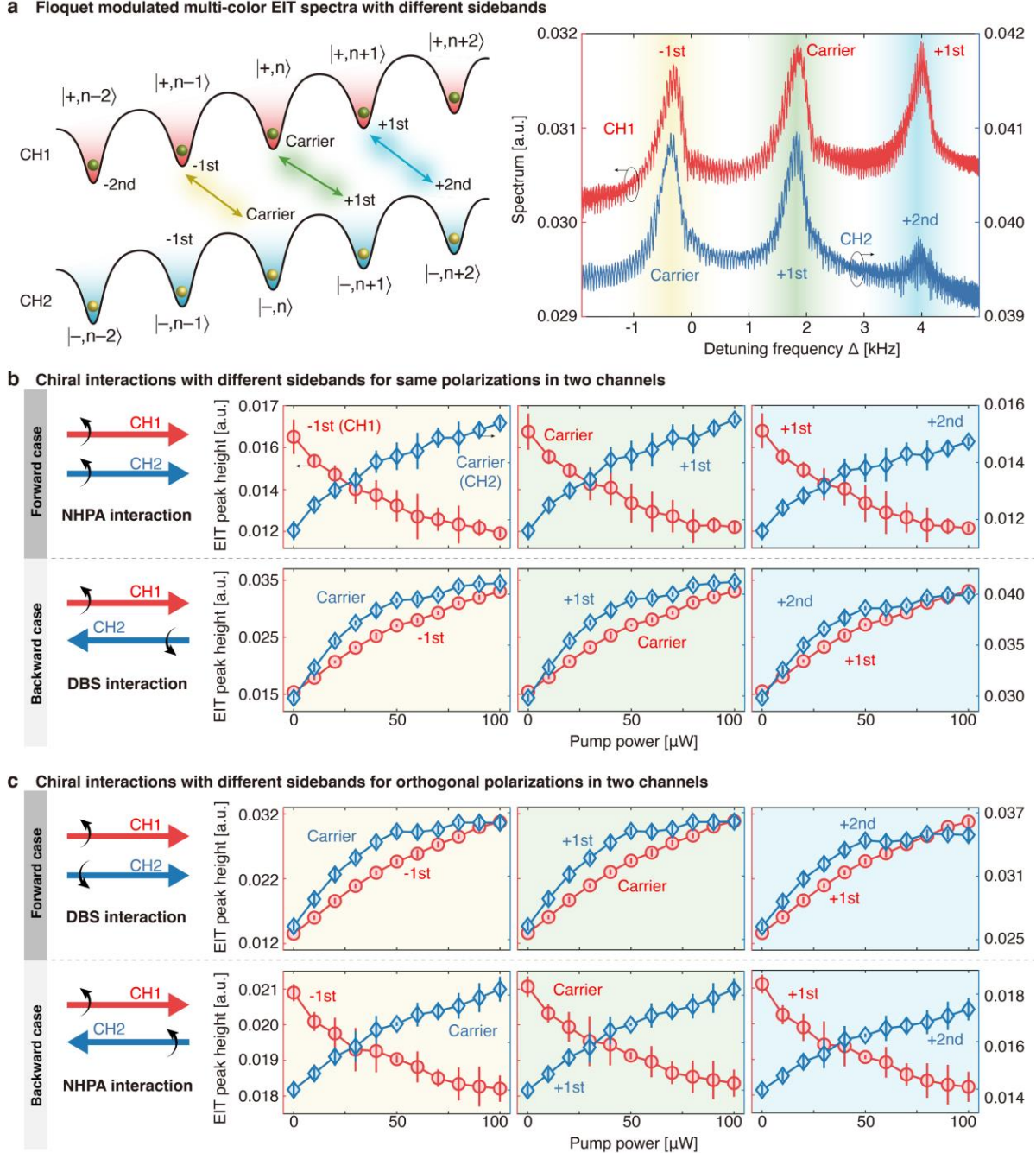

**Extended Data Fig. 5 | Propagation-direction-dependent chiral interactions between different sidebands with Floquet engineering. a**, Floquet modulated multi-color EIT spectra with different sidebands. Compared to the case with same sidebands shown in Fig. 3, the EIT spectra of the light in CH1 (red) and CH2 (blue) show the transparency windows with different sidebands: −1st sideband (CH1) with carrier (CH2), carrier (CH1) with +1st sideband (CH2), and +1st sideband (CH1) with +2nd sideband (CH2). Here, the spectra correspond to the backward case in **b** with pump power at 100 $\mu$W, the Floquet modulation frequency is 2 kHz, and the control frequency in CH2 has a 2 kHz difference from that in CH1. **b**, Nonlinear NHPA and linear DBS interactions between different sidebands are generated in forward and backward cases for two channels with same polarizations, respectively. Such propagation-direction-dependent chiral interactions is revealed by the EIT response amplitudes in the two channels. **c**, Nonlinear NHPA (linear DBS) interaction can also be introduced in backward (forward) case for two channels with orthogonal polarizations.

22